\documentclass{article}

\usepackage[preprint]{neurips_2023}

\usepackage[utf8]{inputenc} 
\usepackage[T1]{fontenc}    
\usepackage{hyperref}       
\usepackage{url}            
\usepackage{booktabs}       
\usepackage{amsfonts}       
\usepackage{nicefrac}       
\usepackage{microtype}      
\usepackage{xcolor}         
\usepackage{tabularx}
\usepackage{booktabs}
\usepackage{rotating}

\title{Identification of cardiovascular diseases through ECG classification using wavelet transformation}

\author{%
    Morteza Maleki \\   
  College of Computing\\
  Georgia Institute of Technology\\
  Atlanta, Georgia\\
  mmaleki3@gatech.edu\\
  \AND
  Foad Haeri\\   
  College of Computing\\
  Georgia Institute of Technology\\
  Atlanta, Georgia\\
  mfhaeri6@gatech.edu\\
}

\begin{document}

\maketitle

\begin{abstract}
Cardiovascular diseases are the leading cause of mortality globally, necessitating advancements in diagnostic techniques. This study explores the application of wavelet transformation for classifying electrocardiogram (ECG) signals to identify various cardiovascular conditions. Utilizing the MIT-BIH Arrhythmia Database, we employed both continuous and discrete wavelet transforms to decompose ECG signals into frequency sub-bands, from which we extracted eight statistical features per band. These features were then used to train and test various classifiers, including K-Nearest Neighbors and Support Vector Machines, among others. The classifiers demonstrated high efficacy, with some achieving an accuracy of up to 96\% on test data, suggesting that wavelet-based feature extraction significantly enhances the prediction of cardiovascular abnormalities in ECG data. The findings advocate for further exploration of wavelet transforms in medical diagnostics to improve automation and accuracy in disease detection. Future work will focus on optimizing feature selection and classifier parameters to refine predictive performance further.
\end{abstract}

\section{Introduction}

Cardiovascular diseases (CVDs) remain the predominant cause of death globally, accounting for an estimated 17.9 million lives each year according to the World Health Organization \cite{WHO2021}. The early diagnosis of CVDs is crucial for effective treatment and management. Electrocardiogram (ECG) tests, which record the electrical activity of the heart over a period, play an essential role in the initial screening and diagnosis of heart conditions.\\ 

However, the manual analysis of ECG data is time-consuming and subject to variability in expert interpretation. As the volume of data generated increases, automated and more accurate methods for ECG analysis become imperative. Recent advances in methodology and machine learning techniques has prompted health researchers to utilize advanced analytics to solve societal and health-related issues, providing pathways for experts to tackle issues impacting population. Machine learning offers promising tools for enhancing the analysis of ECG data, particularly through the application of signal processing techniques.\\

Among various signal processing methods, wavelet transformation has emerged as a powerful technique for analyzing non-stationary physiological signals \cite{Addison2005}. Unlike the Fourier transform, which only provides frequency information, wavelet transforms can capture both time and frequency information, making them particularly suited for the analysis of ECG signals which consist of complex, transient signal components.\\

Despite its potential, the application of wavelet transformation in ECG signal classification is not extensively explored in the literature, especially in the context of machine learning models that can process large datasets efficiently. This research aims to fill this gap by applying wavelet transformation to ECG data for feature extraction and classification, using the MIT-BIH Arrhythmia Database to train and validate various machine learning models. Our study explores both continuous and discrete wavelet transformations to determine their effectiveness in extracting meaningful features from ECG signals that can be used to accurately classify different types of cardiovascular diseases.

\subsection{Literature Review}

The analysis of ECG signals using machine learning techniques has been extensively studied, with various approaches showing promise in enhancing diagnostic accuracy. Early methods predominantly relied on Fourier transforms to analyze the frequency components of ECG signals \cite{Addison2005, MartinsJones2018}. However, the Fourier transform's limitation in capturing time-varying properties of ECG signals often led to inadequate results in real-world scenarios where ECG characteristics fluctuate \cite{HeWu2019}. Wavelet transform, introduced as an alternative to Fourier transform, provides a more nuanced analysis by allowing both time and frequency resolution to adapt dynamically. The adaptability of wavelet transforms makes them particularly suited for ECG signals, which contain non-stationary or transient characteristics that are crucial for identifying cardiovascular abnormalities.\\

Several studies have employed discrete wavelet transforms (DWT) for ECG signal processing. For instance, Addison et al. (2005) demonstrated the potential of DWT in detecting myocardial infarction by extracting detailed features from different ECG signal components \cite{Addison2005}. Their work highlighted the ability of wavelet features to improve the sensitivity and specificity of automated diagnostic systems. Moreover, continuous wavelet transform (CWT) has also been applied to analyze ECG signals for arrhythmia detection. In a comprehensive study by Martins et al. (2018), CWT was used to classify various types of arrhythmias with a high degree of accuracy, using a combination of wavelet-based features and machine learning classifiers \cite{Martins2018CWT}.\\

Recent advances have integrated wavelet transformation with deep learning techniques to further enhance the classification accuracy. Zhang et al. (2020) developed a convolutional neural network model that uses wavelet-transformed ECG signals to automatically detect arrhythmias, achieving state-of-the-art performance \cite{Zhang2020DeepLearning}. This integration of wavelet analysis and deep learning underscores the potential for significant improvements in automated ECG analysis systems. However, despite these advancements, the application of wavelet transforms in ECG classification still presents challenges, particularly in handling large datasets and optimizing feature selection for improved classifier performance. This study aims to address these challenges by exploring both the discrete and continuous wavelet transformations to evaluate their efficacy in feature extraction and subsequent classification accuracy in a large-scale ECG dataset.

\newpage
\section{Material and Methods}

\subsection{Data Sources}
The study utilizes the MIT-BIH Arrhythmia Database \cite{PhysioNetMITDB}, which comprises 48 half-hour excerpts of two-channel ambulatory ECG recordings from 47 subjects studied at the Beth Israel Hospital between 1975 and 1979. The dataset is a well-established resource in the field of cardiovascular disease research and provides a diverse range of ECG signal patterns for analysis. A processed and segmented version of this dataset, prepared for a Kaggle challenge by Kachuee et al., is used, which includes 109,446 samples categorized into five classes. The samples are pre-divided into 87,554 training and 21,892 testing samples, with this study using a 70/30 split of the training dataset for model training and validation \cite{Kachuee2018}.

\subsection{Wavelet Transformation}
Wavelet transformation is utilized to decompose the ECG signals into constituent frequencies and to analyze the non-stationary properties of the ECG data. The transformation enables the isolation of signal components at different scales, which is vital for detecting transient features in ECG signals.\\

\begin{figure*}
\centering
\includegraphics[width=10cm]{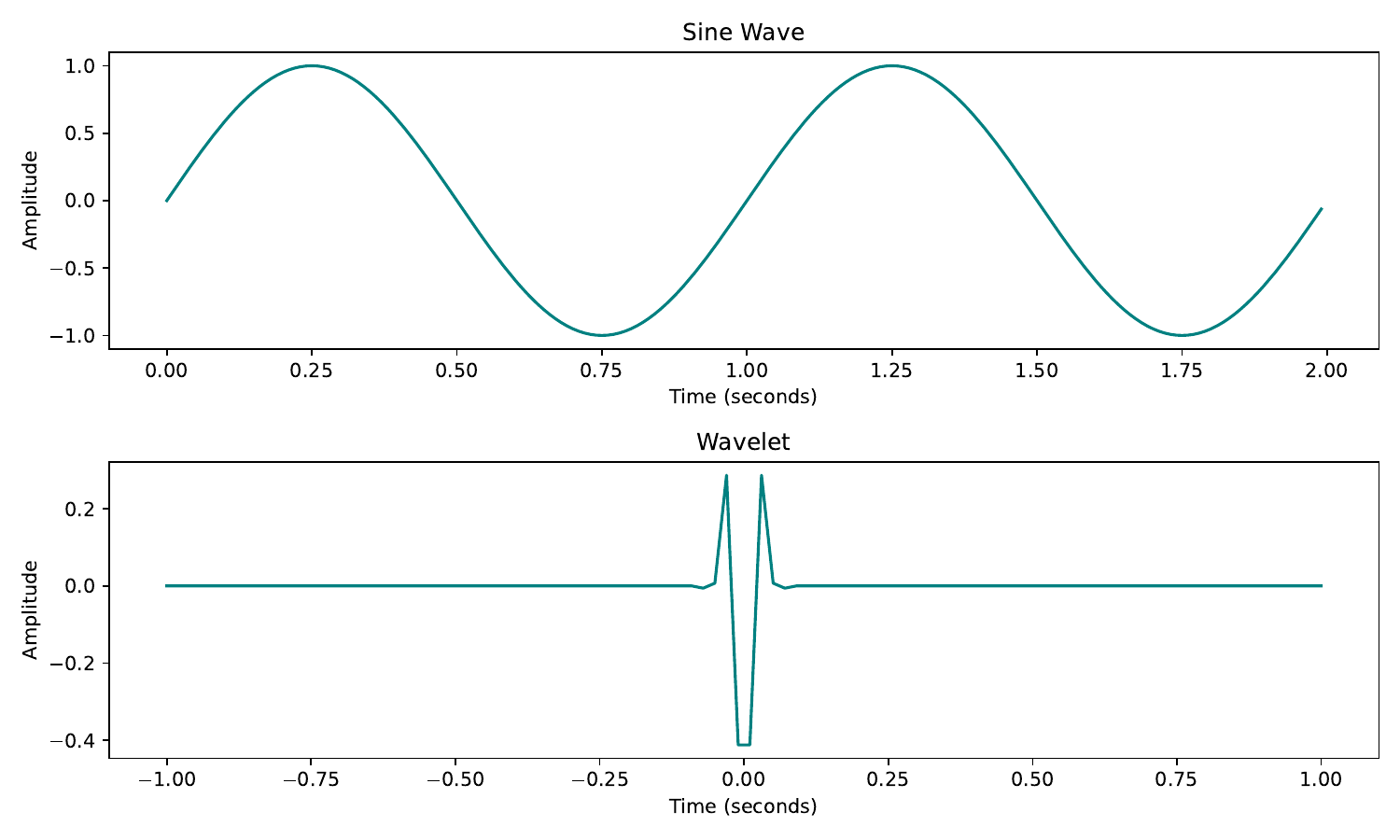}
\caption{Difference between a sine-wave and wavelet used by Fourier and wavelet transform respectively}
\label{figure1}
\end{figure*}

A typical ECG signal consists of a number of underlying oscillatory frequency components superimposed by a noise component. A general signal is a superposition of special scales at different locations in time. Unlike Fourier transform that decomposes a signal into a number of scales using either time or frequency, wavelet transform decomposes a signal into a set of basic functions called wavelets (Figure \ref{figure1}). While Fourier transform uses a linear combination of infinitely-long sine-waves to represent a signal, wavelet transform uses a series of localized wavelets each with different scale, which allows for obtaining time information as well as the frequency information.

\begin{figure*}
\centering

\includegraphics[width=14cm]{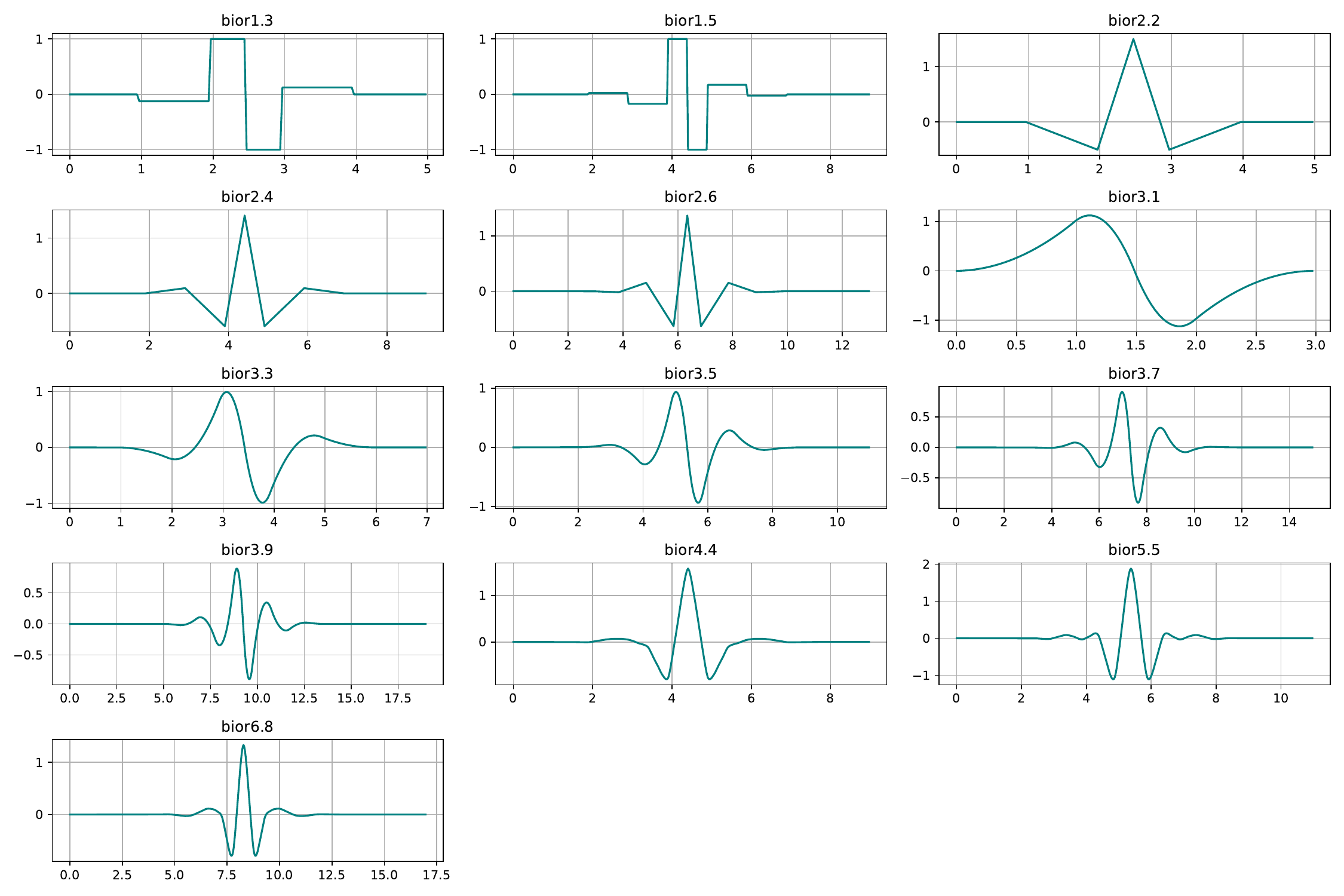}
\caption{ Different wavelets can be applied for decomposition based on the type of a signal }
\label{figure2}
\end{figure*}

Wavelet transformation is done by multiplying the signal with a wavelet at different locations in time and then repeating after a change in scale of the wavelet. The larger scale of a wavelet (longer wavelet) represents a lower frequency and vice versa. Figure \ref{figure2} depicts some members of wavelet families.\\

The CWT is described as follows:\\

\begin{equation}
\label{eq1}
\gamma(a, b) = \int f(t) \Psi_{a,b}(t) \, dt
\end{equation}

which shows how a signal function \( f(t) \) can be decomposed into a set of wavelets \( \Psi_{a,b}(t) \) based on scale and position parameters, \( a \) and \( b \) respectively. The wavelet family members for a signal are generated from a single basic wavelet (mother wavelet) by changing the scale (compressing or stretching) and shifting the position as formulated in Eq. \ref{eq2}:\\

\begin{equation}
\label{eq2}
\Psi_{a,b}(t) = \frac{1}{\sqrt{a}} \Psi\left(\frac{t-b}{a}\right)
\end{equation}

\begin{figure*}
\centering
\includegraphics[width=15cm]{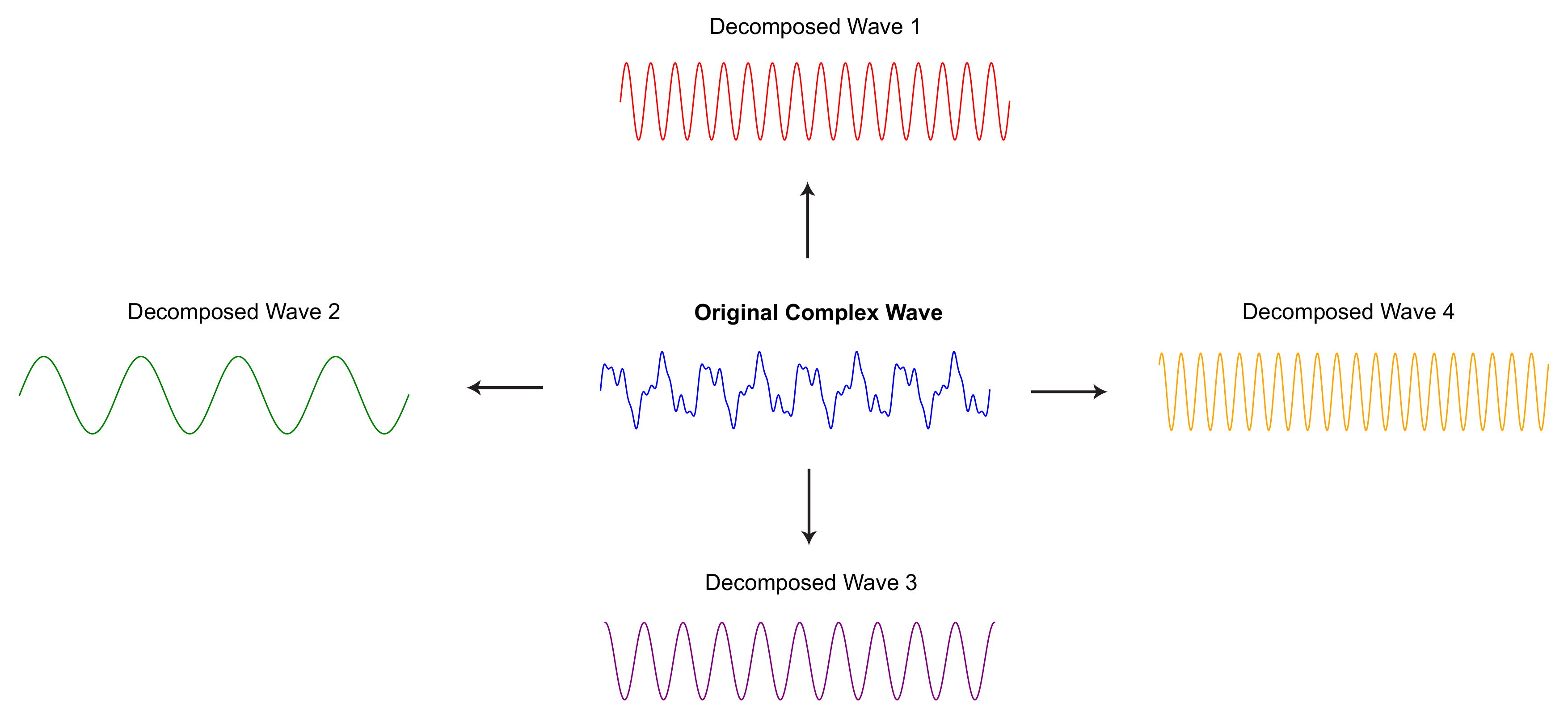}
\caption{Wavelet transformation decomposes a signal into a set of wavelets of different scales and positions}
\label{figure3}
\end{figure*}

This is how a signal is decomposed into constituent wavelets of different scales and positions as described in Figure \ref{figure3}.

\begin{figure*}
\centering
\includegraphics[width=13cm]{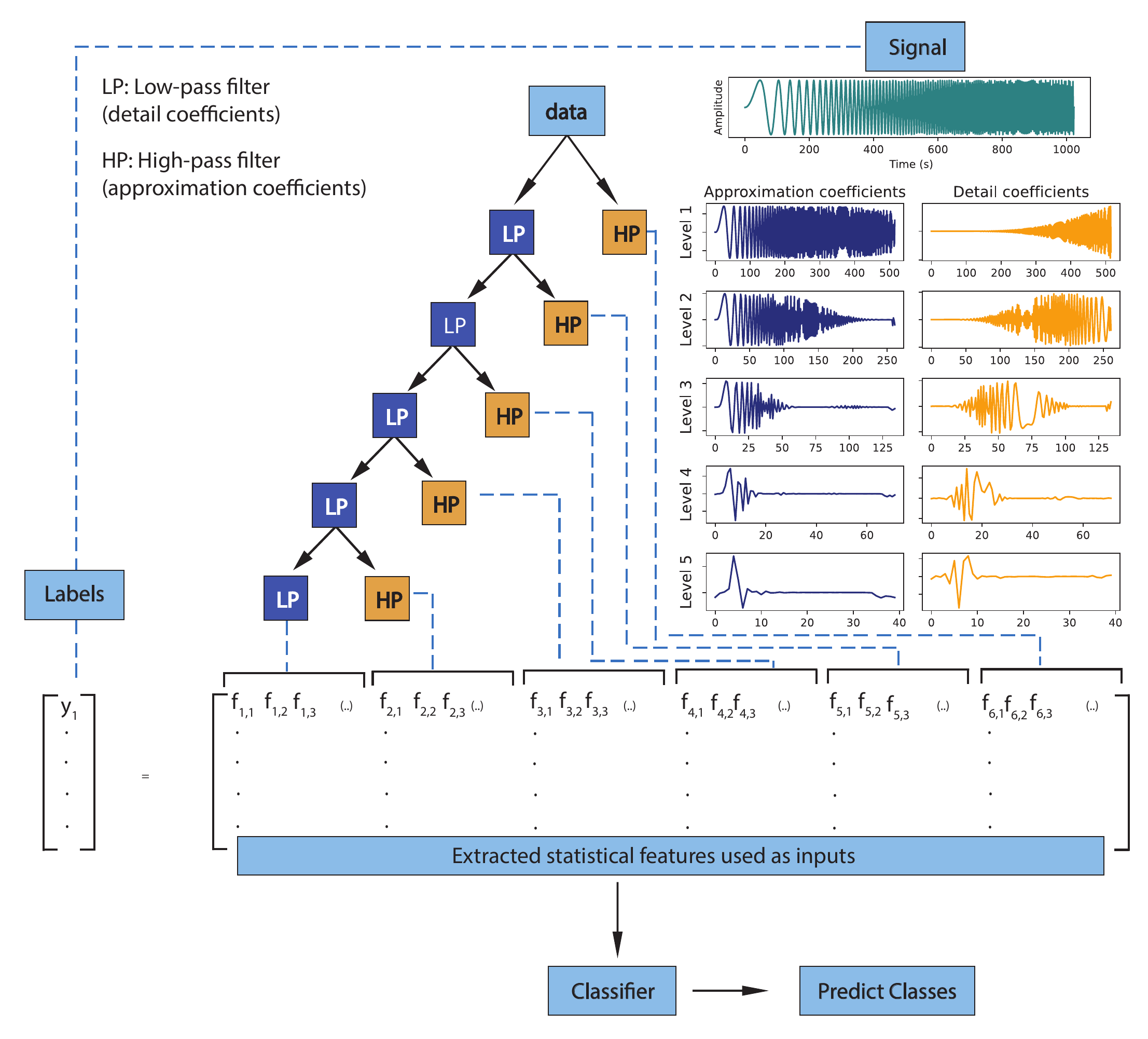}
\caption{Schematic of the classification approach, in which the outputs of DWT function are used to extract statistical features and incorporate them as inputs into the classifier. The original schema is reported in \cite{Taspinar2018} article, and utilized here with some modifications.}
\label{figure4}
\end{figure*} 

To focus on a more manageable set of wavelets produced from an infinite number of wavelets in the continuous wavelet transformation and increase the efficiency, DWT is introduced. Unlike continuous, these wavelets can only be scaled and shifted in discrete steps.\\
 
In practice, DWT is typically implemented to split a signal into a series of frequency sub-bands using high-pass filters (which are returned in the detail coefficients) and low-pass filters (which are returned in the approximation coefficients). The DWT is repeatedly applied on the approximation coefficients of the previous DWT to generate the wavelet transform of the next level. At each level, the original signal is sampled down by a factor of 2 and eventually decomposed to several signals corresponding to different frequency bands as shown in Figure \ref{figure4}.

\subsection{Feature Extraction}

From each level of DWT decomposition, several statistical features are extracted to represent the characteristics of the signal in each frequency band. These features include the mean, median, standard deviation, variance, root mean square value, zero-crossing rate, mean crossing rate, and entropy. The extraction of these features is inspired by the ability of such metrics to capture essential characteristics of the signal relevant to differentiating between normal and abnormal ECG patterns. All extracted features from the sub-bands are then combined into a comprehensive feature vector for each sample.

\subsection{Classification Models}

A range of machine learning classifiers are evaluated for their ability to classify ECG signals based on the extracted features. These classifiers include K-Nearest Neighbors, Support Vector Machines (SVC) with linear and Radial Basis Function (RBF) kernels, Decision Trees, Random Forests, Multi-layer Perceptrons, AdaBoost, Gaussian Naive Bayes (NB), and Gradient Boosting. Each model's performance is assessed based on accuracy, precision, recall, and F1-score, with a focus on optimizing hyperparameters to improve prediction accuracy and model robustness.

\subsection{Implementation Details}

The implementation of the transformation, feature extraction, and classification processes is carried out using Python, leveraging libraries such as PyWavelets for wavelet transformations and scikit-learn for machine learning modeling. The computational work is managed within a Jupyter Notebook environment to ensure reproducibility and facilitate documentation of the code and results. The analysis data and code are provided in the following GitHub repository: \href{https://github.com/mprtrmrtz/wavelet_signaling}{GitHub - Wavelet Signaling}

\section{Results}

The classification of ECG signals involved processing signals that represent both normal and various abnormal heart conditions, including different types of arrhythmias and myocardial infarction. These signals were preprocessed and segmented, with each segment representing a single heartbeat, as depicted in Figure \ref{figure5}.

\subsection{Wavelet Choice and Decomposition}

\begin{figure*}
\centering
\includegraphics[width=10cm]{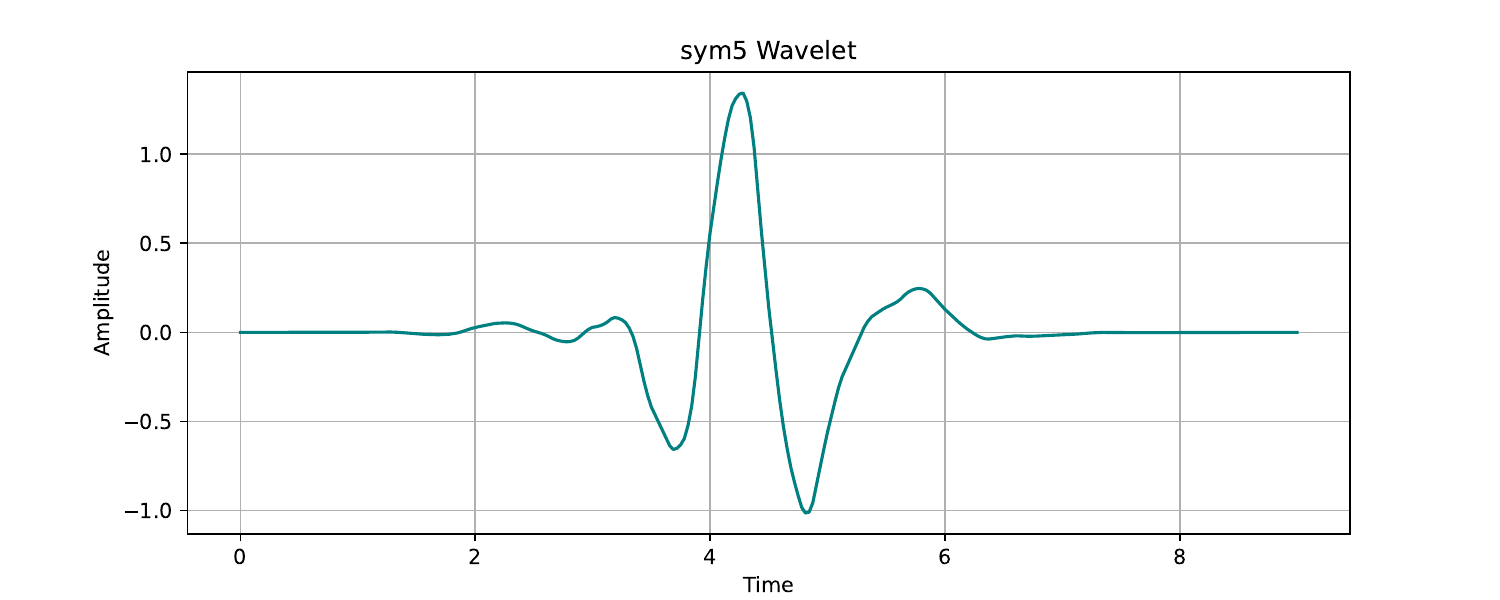}
\caption{Typical shape of sym5 wavelet used to decompose the ECG signals.}
\label{figure6}
\end{figure*}

For signal decomposition, the "sym5" wavelet was chosen due to its nearly symmetric properties and effectiveness in decomposing non-stationary signals like ECG. This choice is crucial as it impacts the quality of the feature extraction process. The sym5 wavelet, a member of the Daubechies family, is known for its balance between time and frequency localization \cite{Mallat1989}. Figure \ref{figure6} illustrates the typical shape of the sym5 wavelet used in this study.

\subsection{Feature Extraction and Dataset Preparation}

Using the DWT, each ECG signal was decomposed into five levels. From each level of decomposition, eight statistical features were extracted, including mean, median, standard deviation, variance, root mean square value, zero-crossing rate, mean crossing rate, and entropy. This resulted in a comprehensive dataset with 40 features per sample, derived from different frequency sub-bands of each decomposed signal. The dataset was then split into a training set (70\%) and a testing set (30\%).

\subsection{Classifier Performance and Evaluation}

\begin{figure*}
\centering
\includegraphics[width=14cm]{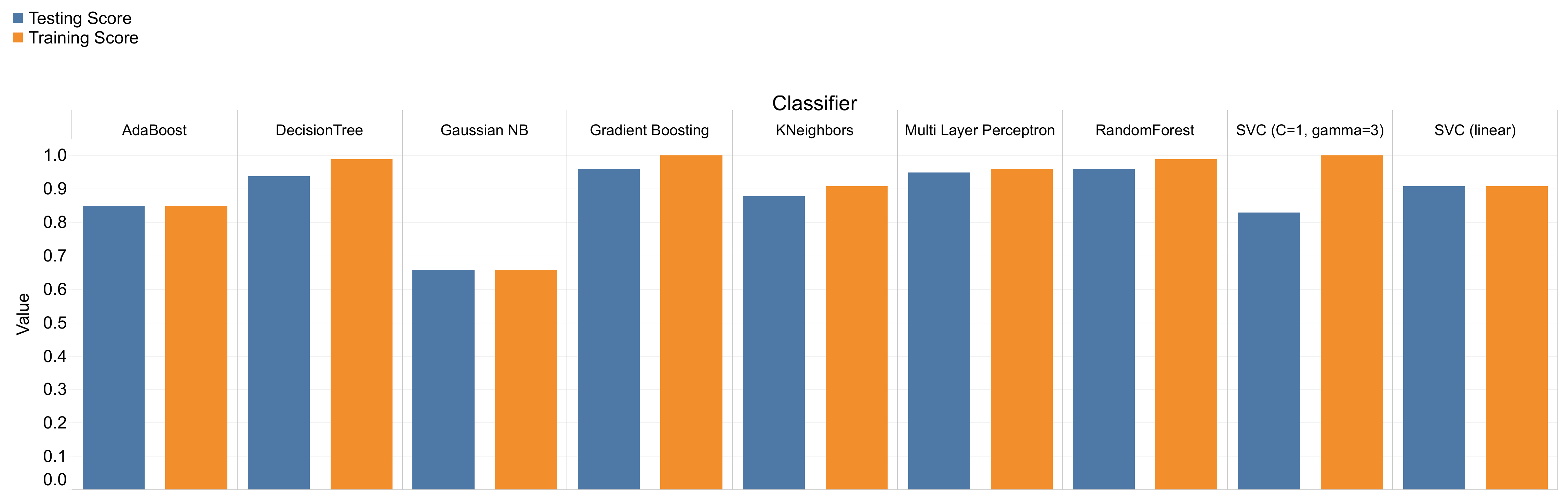}
\caption{Comparison of classifiers’ performances sorted in the order of testing scores}
\label{figure7}
\end{figure*}

The performance of various classifiers was evaluated, as summarized in the Table \ref{table1} and Figure \ref{figure7}, which lists the classifiers along with their hyperparameters, and their respective training and testing accuracies:\\


\begin{table}[h]
\caption{Training and testing accuracy of classifiers used in the study.\label{table1}}
\newcolumntype{C}{>{\centering\arraybackslash}X} 
\begin{tabularx}{\textwidth}{|l|X|c|c|}
\hline
\textbf{Classifier} & \textbf{Hyperparameters} & \textbf{Training Accuracy} & \textbf{Testing Accuracy} \\ \hline
K-Nearest Neighbors & K=5, p=1 & 0.91 & 0.88 \\ \hline
Linear SVC & N/A & 0.91 & 0.91 \\ \hline
RBF SVC & gamma=2, C=1 & 1.00 & 0.83 \\ \hline
Decision Tree & max\_depth=20 & 0.99 & 0.94 \\ \hline
Random Forest & max\_depth=20, max\_features=5, n\_estimates=10 & 0.99 & 0.96 \\ \hline
Multi-Layer Perceptron & hidden\_layer\_sizes = (50,100), alpha = 0.01, max\_iter=1000, activation = ‘tanh’, solver=’adam’ & 0.96 & 0.95 \\ \hline
AdaBoost & N/A & 0.85 & 0.85 \\ \hline
GaussianNB & N/A & 0.66 & 0.66 \\ \hline
Gradient Boost & N\_estimators=10000 & 1.00 & 0.96 \\ \hline
\end{tabularx}
\end{table}


\subsection{Analysis of Classifier Overfitting and Model Complexity}
Classifiers such as Gradient Boost, Decision Tree, and Random Forest demonstrated nearly perfect training accuracy, suggesting potential overfitting issues. These models, along with the RBF SVC, exhibited high training accuracies but varied in testing performance, indicating the need for further hyperparameter tuning and regularization to balance model fit and generalization. Conversely, the GaussianNB and AdaBoost classifiers, with lower accuracies, suggest underfitting, possibly due to the model simplicity or inadequate capture of complex patterns in the data.

\subsection{Significance of Results}

The study confirms the effectiveness of wavelet-based feature extraction in classifying ECG signals. It highlights the critical impact of selecting appropriate wavelets and the depth of feature extraction on the accuracy of different classifiers. The varying performance of the classifiers underscores the importance of careful model selection and hyperparameter optimization in developing robust diagnostic tools.\\

The high performance of the Random Forest and Gradient Boost classifiers, in particular, demonstrates their potential in handling complex patterns and large feature sets derived from wavelet transformations. These results are significant as they suggest that advanced machine learning techniques can significantly enhance the accuracy and reliability of automatic ECG analysis systems, potentially leading to better diagnostic outcomes in clinical settings.\\

Furthermore, the study identifies the challenges of overfitting and underfitting in complex models, providing a valuable foundation for future research. It emphasizes the need for ongoing adjustments and testing in diverse clinical environments to refine these models for practical application. By leveraging the detailed feature sets obtained through wavelet transform, this research paves the way for developing more sensitive and specific ECG classification systems, which are crucial for early detection and management of cardiovascular diseases.\\

\section{Discussion and Conclusions}

The results of this study highlight the potential of wavelet-based feature extraction in enhancing the accuracy of ECG signal classification through machine learning. The use of the sym5 wavelet for signal decomposition leverages its ability to capture both time and frequency information, which is crucial for analyzing the intricate and non-stationary nature of ECG signals. This choice was validated by the high classification accuracies achieved by several of the tested models, particularly the Random Forest and Gradient Boost classifiers, which both demonstrated an excellent balance of high training and testing accuracy.\\

The performance of the Gradient Boost classifier, with perfect training accuracy and very high testing accuracy, suggests that this model was highly effective at learning the distinguishing features within the training dataset and generalizing this learning to new data. However, such a high training accuracy also raises concerns about potential overfitting, even though the high testing accuracy indicates that overfitting was controlled effectively. In contrast, models like the Gaussian Naive Bayes and AdaBoost exhibited lower accuracies, which may indicate that these models are either too simplistic or not adequately configured to capture the complex relationships in the data derived from wavelet-transformed signals.\\

Significant differences in performance among the classifiers also underscore the importance of feature selection and model tuning in achieving optimal results. The dataset's complexity, indicated by 40 distinct features extracted from each ECG signal segment, presents both an opportunity and a challenge. While it provides rich information for classification, it also requires careful handling to avoid dimensionality issues that could obscure meaningful patterns rather than revealing them. This aspect is particularly critical for classifiers that performed suboptimally, such as Gaussian Naive Bayes and AdaBoost, suggesting that further tuning or a different approach to feature handling might be necessary for these models.\\

Moreover, the study's findings contribute to the ongoing discussion about the practical application of machine learning in medical diagnostics. The ability of advanced classifiers to distinguish between different types of cardiac abnormalities with high accuracy holds significant promise for clinical applications, where rapid and accurate diagnosis can drastically improve patient outcomes. Future work could explore the integration of these machine learning models into real-time diagnostic systems, which could aid clinicians in making faster and more accurate decisions.\\

In conclusion, this research not only demonstrates the efficacy of wavelet-based feature extraction and machine learning in classifying ECG signals but also highlights critical areas for future investigation, including model overfitting, the impact of feature selection, and the translational potential of these technologies in clinical settings. The exploration of these areas will be crucial for moving from theoretical models to practical applications that can benefit patient care in real-world scenarios.

\bibliographystyle{plainnat}
\bibliography{references}

\end{document}